\documentclass[prl,twocolumn,superscriptaddress,showpacs]{revtex4}
\usepackage{epsfig}

\begin{document}

\title{Nature of $e_g$ Electron Order in La$_{1-x}$Sr$_{1+x}$MnO$_4$ }

\author{S. Larochelle} 
\affiliation{Department of Physics, Stanford University, Stanford, CA 94305} 
\author{A. Mehta} 
\author{N. Kaneko}
\affiliation{Stanford Synchrotron Radiation Laboratory, 
    Stanford Linear Accelerator Center, Stanford, CA 94309}
\author{P. K. Mang} 
\author{A. F. Panchula} 
\affiliation{Department of Applied Physics, Stanford University, Stanford,
    CA 94305}
\author{L. Zhou}
\affiliation{T. H. Geballe Laboratory for Advanced Materials,
    Stanford University, Stanford, CA 94305} 
\author{J. Arthur}
\affiliation{Stanford Synchrotron Radiation Laboratory,
    Stanford Linear Accelerator Center, Stanford, CA 94309} 
\author{M. Greven} 
\affiliation{Department of Applied Physics, Stanford University, Stanford,
    CA 94305} 
\affiliation{Stanford Synchrotron Radiation Laboratory, 
    Stanford Linear Accelerator Center, Stanford, CA 94309}

\begin{abstract}

Synchrotron x-ray scattering measurements of the low-temperature structure
of the single-layer manganese oxide La$_{1-x}$Sr$_{1+x}$MnO$_4$, over the
doping range $0.33 \le x \le 0.67$, indicate the existence of three distinct
regions: a disordered phase ($x \le 0.4$), a charge-ordered phase
($x \ge 0.5$), and a mixed phase ($0.4 < x < 0.5$). For $x>0.5$, the
modulation vector associated with the charge order is incommensurate with
the lattice and depends linearly on the concentration of $e_g$ electrons.
The primary superlattice reflections are strongly suppressed along the
modulation direction and the higher harmonics are weak, implying the
existence of a largely transverse and nearly sinusoidal structural
distortion, consistent with a charge-density wave of the $e_g$ electrons.

\end{abstract}


\maketitle

Perovskite-derived transition metal oxides have attracted much attention
over the past decade because of their unusual electronic properties.
Strong electron correlations give rise to such phenomena as high-temperature
superconductivity in layered cuprates and stripe-like order in layered
cuprates and nickelates. In the case of the manganites, an additional strong
coupling between charge and lattice degrees of freedom leads to a very
rich electronic phase diagram in which crystallographic and magnetic
structures as well as transport properties are intimately related. The
competition among various phases has been closely associated \cite{Moreo99}
with the colossal magnetoresistance (CMR) observed in these materials. For
example, the perovskite La$_{1-x}$Ca$_{x}$MnO$_3$ exhibits two low-temperature
phases at doping $x \sim 0.5$, one a ferromagnetic metal
and the other an antiferromagnetic
insulator \cite{Mori98a,Huang00}. The insulating phase at the doping level
$x=0.5$, where there is on average half an $e_g$ electron per manganese
atom, is characterized by a complex checkerboard arrangement of ordered $e_g$
electrons, orbitals and spins
\cite{Wollan55,Goodenough55,Tokura00}. The ordered phase of
La$_{1-x}$Ca$_{x}$MnO$_3$ extends to rather high doping where the average
number $n_e = 1 - x$ of $e_g$ electrons decreases markedly and, consequently,
the low-temperature unit cell becomes very large \cite{Chen97,Mori98b}.

CMR has been observed in the perovskite and double-layer manganites, but
not in the single-layer system \cite{Moritomo95,Bao96}, the most nearly
two-dimensional member of this series. Nevertheless, there are signs that
the physics of La$_{1-x}$Sr$_{1+x}$MnO$_4$ is similar to that of the
perovskite
manganites. For example, similar antiferromagnetic order has been reported
for La$_{0.50}$Sr$_{1.50}$MnO$_4$ \cite{Sternlieb96}. Moreover, the
low-temperature phase is sensitive to magnetic fields \cite{Tokunaga99b}, and
it extends to high doping (low $e_g$ electron concentration) \cite{Bao96}. A
systematic study of the low-temperature structural phases of
La$_{1-x}$Sr$_{1+x}$MnO$_4$ should provide valuable insights into the effect
of dimensionality on the properties of the manganites. Such a study may also
contribute to a deeper understanding of the single-layer transition metal
oxides.

In this Letter, we report a non-resonant x-ray scattering study of
La$_{1-x}$Sr$_{1+x}$MnO$_4$ ($0.33 \le x \le 0.67$). We first
consider the low-temperature structure of
La$_{0.50}$Sr$_{1.50}$MnO$_4$. Our data provide a more complete
picture than previous neutron [11] and x-ray [13] scattering
experiments. We then extend our investigation to study the effects
of varying the $e_g$ electron concentration in the MnO$_2$ layers.
We find three distinct regions: disordered ($x \le 0.4$),
mixed-phase ($0.4 < x < 0.5$), and charge-ordered ($x \ge 0.5$).
Above $x=0.5$, the ordering of $e_g$ electrons results in a
structural distortion whose modulation period only depends on
$n_e$. The distortion is largely transverse and nearly sinusoidal,
consistent with charge-density-wave order of the $e_g$ electrons.

Samples of La$_{1-x}$Sr$_{1+x}$MnO$_4$ were prepared from stoichiometric
amounts of La$_2$O$_3$, MnO$_2$ and SrCO$_3$ powders. The mixtures were
calcinated three times for twelve hours between 1300$^\circ$C and
1360$^\circ$C. The calcinated powders were then pressed into rods of 5mm
diameter and sintered at 1600$^\circ$C for twelve hours. Single crystals
were grown by the floating-zone method at a rate of 6mm/hour in an oxygen
atmosphere of 5bar.
Crystal pieces of dimensions 2x4x1mm$^3$ were cut from the grown boules and
were mounted inside a closed-cycle refrigerator.
The mosaic widths of the crystals used in this study
were $0.02-0.06^\circ$ full width at half maximum (FWHM).
The data were collected
using a four-circle diffractometer at beamline 7-2 of the Stanford Synchrotron
Radiation Laboratory. A monochromatic x-ray beam was obtained from the wiggler
spectrum via a Si (111) double-crystal monochromator. 
In order to study the bulk structure, an x-ray energy of 14keV
was selected, providing a penetration depth of about $40\mu m$.

The low-temperature phase of La$_{0.50}$Sr$_{1.50}$MnO$_4$ was
probed extensively via scans along the high-symmetry directions in
reciprocal space. Two such scans are presented in Fig. 1. The
essential features of the ordered structure do not change below
the charge-ordering temperature $T_{CO} \approx 240$K. In addition
to the peaks present in the high-temperature (I{\footnotesize
4/mmm}) structural phase (henceforth referred to as high-symmetry
peaks), peaks with wavevectors $(\frac{1}{4},\frac{1}{4},0)_t$ and
$(\frac{1}{2},\frac{1}{2},0)_t$ were found (the subscripts $t$ and
$o$ indicate, respectively, the tetragonal and orthorhombic unit
cells). The $(\frac{1}{4},\frac{1}{4},0)_t$ peaks ($h$-odd peaks
in the orthorhombic notation used in Fig. 1) suffer a near
extinction along $(h,h,0)_t$ ($(h,0,0)_o$ in orthorhombic
notation). The presence of a quarter-wave modulation has been
reported in electron diffraction measurements
\cite{Moritomo95,Bao96}. A neutron diffraction study found only a
half-wave modulation and concluded that the quarter-wave
modulation seen in electron diffraction comes from a surface phase
\cite{Sternlieb96}. The measurements reported here were performed
at relatively high x-ray energy (14keV) to achieve a bulk
penetration depth and were reproduced on samples from three
different growths. Moreover, preliminary neutron diffraction 
on a larger sample with mosaic width of about $0.2^{\circ}$ (FWHM)
has confirmed the presence of the quarter-wave peaks
in the bulk.

The existence of a
quarter-modulation wavevector along $(h,h,0)_t$ seems to imply that
the $a$ and $b$ axes of the low-temperature phase are quadrupled. However,
a smaller (half as large) orthorhombic unit cell, rotated by 45$^{\circ}$
relative to the high-symmetry (I{\footnotesize 4/mmm}) cell, and
its 90$^{\circ}$ twin, are adequate to index all the observed reflections:
$a_o \approx 2 \sqrt{2} a_t$, 
$b_o \approx \sqrt{2}a_t$, and  $c_o = c_t$ ($a_t$ and
$c_t$ are the tetragonal lattice constants). This orthorhombic unit cell 
is consistent with the one used to describe the charge-ordered phases 
of the perovskite La$_{0.50}$Ca$_{0.50}$MnO$_3$ \cite{Radaelli97} and 
the bi-layer material LaSr$_{2}$Mn$_2$O$_7$ \cite{Argyriou00}. However, 
it is different from the cell proposed by Sternlieb {\it et al.}
for La$_{0.50}$Sr$_{1.50}$MnO$_4$ \cite{Sternlieb96}.

\begin{figure}
\epsfig{file=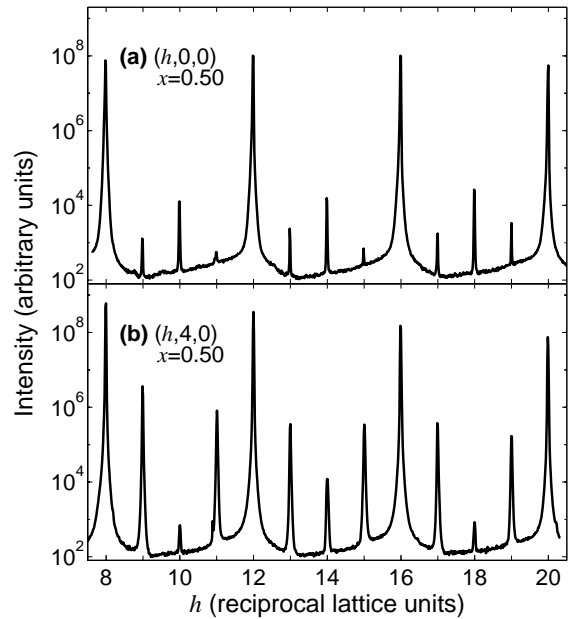,width=7.5cm}
\caption{X-ray diffraction scans of La$_{0.50}$Sr$_{1.50}$MnO$_4$ in the
    low-temperature phase ($T=7$K) along (a) $(h,0,0)_o$ and (b)
    $(h,4,0)_o$ (``$o$'' indicates the orthorhombic unit cell). The peaks at
    $h = 8, 12, 16$ and 20 are present in the high-temperature phase. The
    superlattice peaks at odd-integer values of $h$ are strongly suppressed
    along $(h,0,0)_o$. Note the logarithmic intensity scale. }
\end{figure}

After an extensive survey of reciprocal space we are forced
to conclude that the low-temperature structural symmetry is lower than
previously reported.  It is no higher than the orthorhombic extinction class
A$---$.
A common
characteristic of these structures is that they contain three
unique Mn sites in a $1:1:2$ ratio and that equivalent Mn sites
are located as far apart from each other as possible.
Since some of the quarter-wave peaks are very intense (up to $\sim
1\%$ of the intensity of the high-symmetry peaks, see Fig. 1) and
are visible in neutron as well as in non-resonant x-ray
scattering, the predominant contribution to the intensity of these
peaks must come from a structural distortion and not from charge
or orbital ordering on the Mn sites. These $h$-odd reflections
are, however, heavily suppressed along $(h,0,0)_o$, implying that
the structural distortions are not isotropic, but mostly
orthogonal to the modulation direction $h$. Such a structural
distortion is probably a shear-type distortion of the Mn-O
octahedra (or a Jahn-Teller distortion) similar to the one
proposed for La$_{0.50}$Ca$_{0.50}$MnO$_3$ \cite{Radaelli97} and
LaSr$_{2}$Mn$_2$O$_7$ \cite{Argyriou00}, rather than a
breathing-type distortion of the oxygen octahedra as proposed in
Ref. \cite{Sternlieb96}. In light of this, we note that the
resonant scattering signal at the quater-wave positions, observed
by Murakami {\it et al.} \cite{Murakami98} for $x=\frac{1}{2}$,
either could be ascribed to orbital ordering or to the charge
asymmetry around the Mn ions associated with the structural distortion
reported here \cite{Benfatto99}. To fully characterize the 
low-temperature distortion and the role charge and orbital order play
in manganites, comprehensive and accurate high-resolution
crystallography will be necessary.

\begin{figure}
\epsfig{file=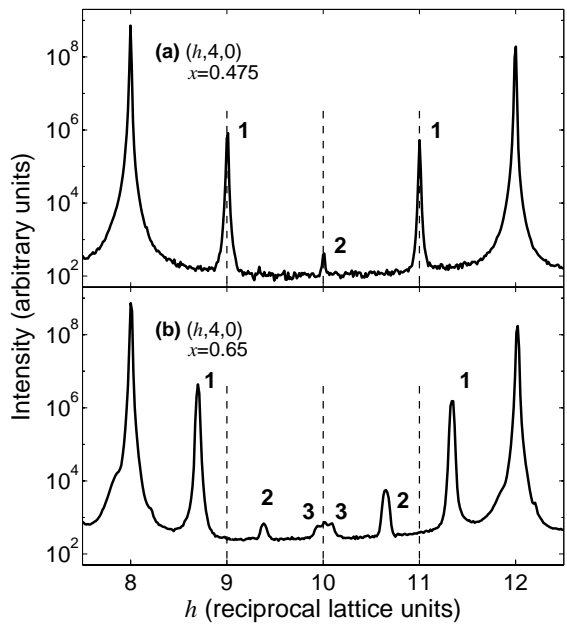,width=7.5cm}
\caption{X-ray diffraction scans of La$_{1-x}$Sr$_{1+x}$MnO$_4$ in the
    low-temperature phase ($T=7$K) along $(h,4,0)_o$ for (a) $x=0.475$
    and (b) $x=0.65$. The vertical dashed lines indicate the commensurate
    positions. 1, 2 and 3 label, respectively, the first, second and third
    harmonics of the low-temperature distortion. Note the logarithmic
    intensity scale. }
\end{figure}

The nature of the distortion can be more clearly identified by
studying the effects of varying the $e_g$ electron concentration
$n_e = 1-x$. We observed superlattice peaks in all samples with
$x>0.40$, up to $x=0.67$, the highest value investigated (see Fig. 2).
For $x > 0.50$, the superlattice modulation vector changes linearly with
$n_e$, as shown in Fig. 3(a). For $x=0.50$, the superlattice
modulation doubles the high-temperature structure (along the
tetragonal base diagonal), and for $x=0.67$, it triples it. This
linear dependence of the wave vector is similar to that observed
in La$_{1-x}$Ca$_{x}$MnO$_3$ for $x>0.5$ \cite{Chen97} and, in
particular, at $x=\frac{2}{3}$
\cite{Mori98b,Radaelli99,Fernandez99,Wang00}. While at
$x=\frac{1}{2}$ and $x=\frac{2}{3}$ commensurate wavevectors are
observed, the ordering is best understood, at all doping levels
$\frac{1}{2} \leq x \leq \frac{2}{3}$, as a modulation whose
period only is a function of $n_e$. Because the superlattice
modulation is directly correlated with $n_e$, it is likely that
the structural phase transition is driven by the ordering of the
$e_g$ electrons.

Second and third diffraction harmonics, much weaker than the
primary (labeled as 2, 3 and 1 in Fig. 2), are visible at $(\pm
2\epsilon,0,0)_o$ and $(\pm 3\epsilon,0,0)_o$. The relative
weakness of the higher
harmonics suggests that the structural distortion is essentially
sinusoidal \cite{Axe80}. The widths of the superlattice peaks
are comparable to those of the high-symmetry peaks which implies
that the low-temperature phase exhibits long-range order.

\begin{figure}
\epsfig{file=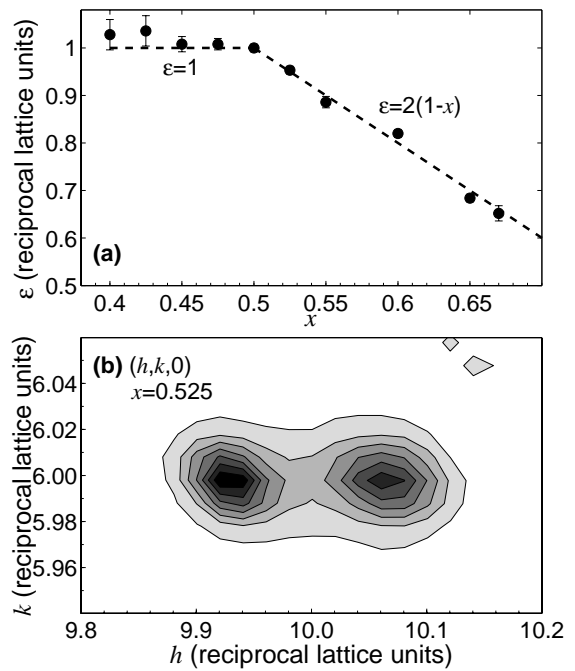,width=7.5cm} 
\caption{(a) Superlattice wave vector $(\pm\epsilon,0,0)_o$ as a
function of
  $x$ for La$_{1-x}$Sr$_{1+x}$MnO$_4$. The dashed line for $x>0.5$ is
  $\epsilon = 2(1-x) = 2n_e$. (b) Linear-scale contour map (10\% contours)
  of the scattering intensity around (10,6,0)$_o$ for $x=0.525$.}
\end{figure}

The nearly sinusoidal structural distortion, together with the
linear variation of the modulation wavevector with doping,
precludes any model in which the $e_g$ electron order is too
closely linked to the underlying cationic lattice, such as the
bi-stripe model of Ref.\cite{Mori98b} or the
discommensurate-stripe model proposed for the single-layer
nickelates \cite{Yoshizawa00}. A better description is given by a
nearly sinusoidal structural distortion, probably associated with
a charge-density wave. The variation of the charge density, with
equivalent Mn sites located as far apart as possible, is similar
to the ``Wigner-crystal'' arrangement of the
$e_g$ electrons proposed for La$_{0.333}$Ca$_{0.667}$MnO$_3$
\cite{Radaelli99}.

A $2^\circ$ rotation of the modulation direction 
with respect to the high-temperature lattice was found from electron
diffraction of La$_{0.33}$Ca$_{0.67}$MnO$_3$
\cite{Wang00}. We did not observe any rotation of the structural
modulation in La$_{1-x}$Sr$_{1+x}$MnO$_4$.
This is shown for $x=0.525$ 
in Figure 3(b).

For $x \le 0.40$, there exists no superlattice structure at low
temperature. Rather, we observe very weak diffuse scattering
similar to what we find for the disordered 
phase of $x \ge 0.50$ above the
charge-ordering temperature. 
This diffuse intensity may arrise from short-range polaron-polaron
correlations similar to those reported 
for the paramagnetic (insulating) phase of
La$_{0.7}$Ca$_{0.3}$MnO$_3$ \cite{Adams00}.
Thus, there is no
long-range distortion of the high-temperature structure for 
$x \le 0.40$, 
and the $e_g$ electrons are not ordered. Unlike the CMR
manganites, La$_{1-x}$Sr$_{1+x}$MnO$_4$ remains insulating in this
region of doping \cite{Moritomo95,Bao96}.

\begin{figure}
\epsfig{file=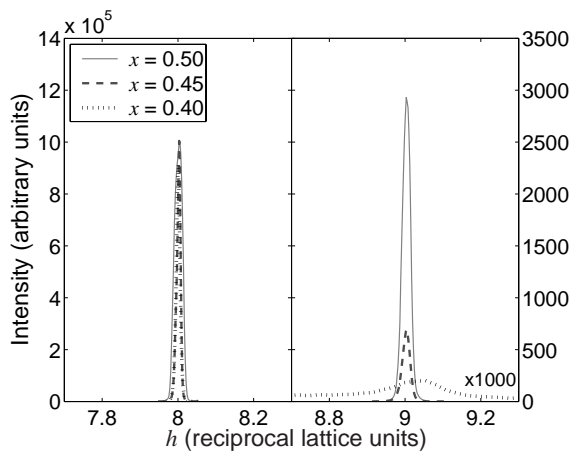,width=7.5cm}
\caption{H-scans through the $(8,4,0)_o$ and $(9,4,0)_o$ reflections for
    $x = 0.40$, 0.45 and 0.50 ($T=100$K). The $(8,4,0)_o$ peak intensities
    are normalized to 10$^6$. Below $x=0.50$, the $(9,4,0)_o$ superlattice
    peak intensity decreases considerably with decreasing $x$.
    For $x=0.40$, the peak is noticeably broadened and its intensity is
    $\sim 10^4$ weaker than for $x=0.50$.}
\end{figure}

For $0.40<x<0.50$, the modulation vector remains the same as for
$x=0.50$. However, as can be seen from Fig. 4, the intensity of
the superlattice peak increases as $x$ increases toward $x=0.50$.
The simplest explanation for this behavior is that the material
separates into charge-ordered regions of 0.5 $e_g$ electrons per
Mn site and disordered regions of approximately 0.6 $e_g$
electrons per Mn site. Additional support for this interpretation
comes from the widths of the peaks. For $x \ge 0.45$, the widths
of the superlattice peaks are nearly constant and comparable to
the widths of the high-symmetry peaks. However, for $x<0.45$, the
superlattice peak widths are ten to twenty times broader and the
peak intensities are much reduced, implying that the remaining
ordered domains are of relatively small size.

In conclusion, we have used synchrotron x-ray scattering to
investigate the nature of the low-temperature structural
distortion of La$_{1-x}$Sr$_{1+x}$MnO$_4$ and its dependence on
the $e_g$ electron concentration $n_e = 1 - x$. At $x=0.50$, our
measurements indicate a shear-type structural distortion of the
Mn-O octahedra. As a function of doping $x$, we found three
distinct regions. For $x \ge 0.50$, the largely transverse
structural distortion exhibits long-range order and the modulation
vector of the distortion varies linearly with $n_e$: $\epsilon = 2
n_e$. This distortion, which is similar to that found for $x=2/3$
in La$_{1-x}$Ca$_x$MnO$_3$ \cite{Radaelli99,Wang00}, appears to be
associated with a nearly sinusoidal charge-density wave. Samples
with $x \le 0.40$ exhibit no long-range
superstructural order, only diffuse scattering peaks, perhaps due
to short-range polaron-polaron correlations. The intermediate
region, $0.40 < x < 0.50$, is best understood as a mixture of
the ordered and the disordered phases. This phase diagram is
reminiscent of that of the CMR perovskite La$_{1-x}$Ca$_x$MnO$_3$.
The structure of the ordered phase appears to primarily depend on
$n_e$, and to be rather insensitive to the dimensionality of the
lattice. Further, both the single-layer system and the CMR
perovskites phase separate for $n_e>0.5$. Nevertheless, in this 
regime their macroscopic low-temperature transport and
magnetic properties differ dramatically.

We would like to acknowledge helpful discussions with B. W.
Batterman, F. Bridges, E. Dagatto, S. Ishihara, S. A. Kivelson, J.
W. Lynn, and S. Maekawa. SSRL is operated by Stanford University
for the U.S. Department 
of Energy, Office of Basic Energy Sciences. 
This work was supported by the U.S. Department of Energy under Contract
Nos. DE-FG03-99ER45773-A001 and DE-AC03-76SF00515,
and by NSF Grant No. DMR9400372. M.G. was also supported 
by the A.P. Sloan Foundation.

\end{document}